# RADIA: RNA and DNA Integrated Analysis for Somatic Mutation Detection

Amie J. Radenbaugh[1], Singer Ma[1], Adam Ewing[1], Joshua Stuart[1], Eric Collisson[2], Jingchun Zhu[1], David Haussler[1,3]

[1] Center for Biomolecular Science and Engineering, 1156 High Street, University of California, Santa Cruz, CA 95064
[2] Division of Hematology/Oncology, 1600 Divisadero Street, University of California, San Francisco, CA 94143
[3] Howard Hughes Medical Institute

## Abstract

The detection of somatic single nucleotide variants is a crucial component to the characterization of the cancer genome. Mutation calling algorithms thus far have focused on comparing the normal and tumor genomes from the same individual. In recent years, it has become routine for projects like The Cancer Genome Atlas (TCGA) to also sequence the tumor RNA. Here we present RADIA (RNA and DNA Integrated Analysis), a method that combines the patient-matched normal and tumor DNA with the tumor RNA to detect somatic mutations. The inclusion of the RNA increases the power to detect somatic mutations, especially at low DNA allelic frequencies. By integrating the DNA and RNA, we are able to rescue back calls that would be missed by traditional mutation calling algorithms that only examine the DNA.

RADIA was developed for the identification of somatic mutations using both DNA and RNA from the same individual. We demonstrate high sensitivity (84%) and very high specificity (98% and 99%) in real data from endometrial carcinoma and lung adenocarcinoma from TCGA. Mutations with both high DNA and RNA read support have the highest validation rate of over 99%. We also introduce a simulation package that spikes in artificial mutations to real data, rather than simulating sequencing data from a reference genome. We evaluate sensitivity on the simulation data and demonstrate our ability to rescue back calls at low DNA allelic frequencies by including the RNA. Finally, we highlight mutations in important cancer genes that were rescued back due to the incorporation of the RNA.

# 1 Introduction

Much of our current understanding of cancer has come from investigating how normal cells are transformed into cancerous cells through the stepwise acquisition of somatic genomic abnormalities. These events include point mutations, insertions and deletions (INDELs), chromosomal rearrangements, and changes to the copy number of segments of DNA. Transforming a normal human cell into a malignant, immortal cancer cell line requires an estimated five to seven genetic alterations in key genes and pathways (Hahn *et al.*, 1999; Hanahan and Weinberg, 2000). Not surprisingly, much research has been devoted to determining how cancer cells are able to acquire their abilities through the accumulation of somatic mutations.

The Cancer Genome Atlas (TCGA) project has produced exome-wide data from thousands of tumors and patient-matched normal tissues. With the development of RNA Sequencing (RNA-Seq), TCGA began providing an additional high-throughput tumor sequence dataset. These three datasets consisting of tumor and patient-matched normal DNA, and tumor RNA have become a new standard in cancer genomics. RNA-Seq enables one to investigate the consequences of genomic changes in the RNA transcripts they encode to better characterize 1) germline variants, 2) somatic mutations, and 3) variants in the RNA that are not found in the DNA that could be the result of RNA editing (Gott and Emeson, 2000).

Over the next few years, many more whole-genome and exome-capture DNA and RNA-Seq BAM (the binary version of Sequence Alignment/Map (Li *et al.*, 2009)) files will become available. TCGA has collected up to 10,000 tissue samples from more than 20 types of cancer. There is a clear need for an efficient method for the combined analysis of patient-matched tumor DNA, normal DNA, and tumor RNA. Here we present a method called RADIA to identify and characterize alterations in cancer using DNA and RNA obtained by high-throughput sequencing data.

Somatic mutation calling is traditionally performed on patient-matched pairs of tumor and normal genomes/exomes. The ability to accurately detect somatic mutations is hindered by both biological and technical artifacts that make it difficult to obtain both high sensitivity and high specificity. Different mutation calling algorithms often disagree about putative mutations in the same source data, and frequently have discernible systematic differences due to the trade-off between sensitivity and specificity (Roberts *et al.*, 2013). This is especially true for somatic mutations with low variant allele frequencies (VAFs). By creating an algorithm that utilizes both DNA and RNA, we have increased the power to detect somatic mutations, especially at low variant allele frequencies.

RADIA combines patient-matched tumor and normal DNA with the tumor RNA to detect somatic mutations. The DNA Only Method (DOM) (Fig. 1) uses just the tumor/normal pairs of DNA (ignoring the RNA), while the Triple BAM Method (TBM) (Fig. 1) uses all three datasets from the same patient to detect somatic mutations. The mutations from the TBM are further categorized into 2 sub-groups: RNA Confirmation and RNA Rescue calls (Supplementary Fig. 1). RNA Confirmation calls are those that are made by both the DOM and the TBM due to the strong read support in both the DNA and RNA. RNA Rescue calls are those that had very little DNA support, hence not called by the DOM, but strong RNA support, and thus called by the TBM. RNA Rescue calls are typically missed by traditional methods that only interrogate the DNA.

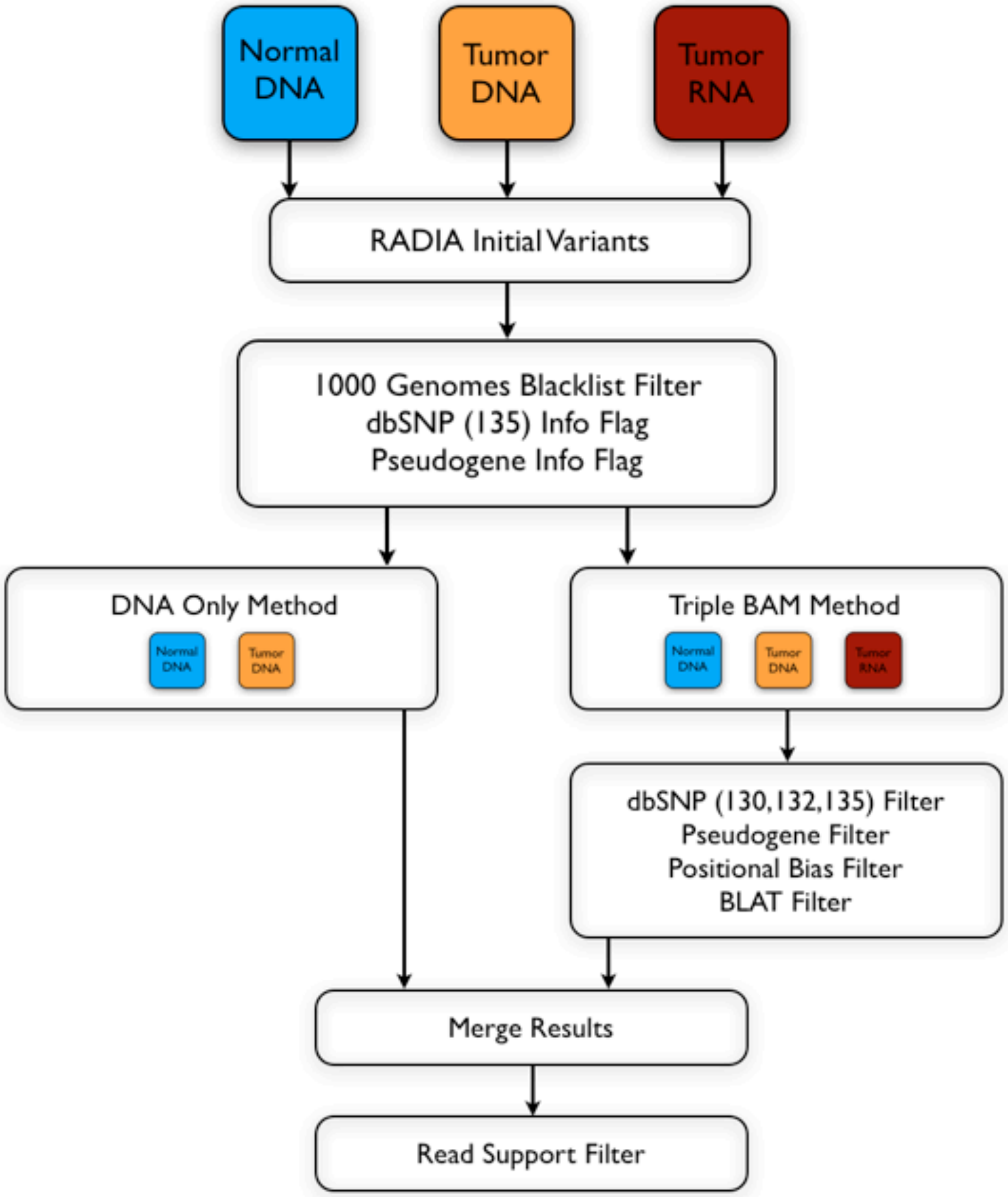

**Fig. 1. Overview of the RADIA work-flow for identifying somatic mutations.** The normal DNA, tumor DNA, and tumor RNA BAMs are processed in parallel and initial low-level variants are identified. The variants are filtered by the DNA Only Method using the pairs of normal and tumor DNA and by the Triple BAM Method using all three datasets. The calls from the two methods are merged and output in VCF format.

We have applied RADIA to data derived from nearly 1,900 patients representing seven different cancer types from TCGA (Supplementary Table 1). Overall, the RNA Rescue calls that are made possible by the incorporation of the RNA-Seq data provide a 2-7% increase in somatic mutation calls compared to the DOM (Supplementary Table 1). Many of these mutations were new discoveries that were not previously found by other mutation callers in TCGA. Of these new discoveries, some mutations were found in well-known cancer genes that were heavily mutated in a specific cohort. We also find mutations in new samples where the same gene has already been identified as harboring mutations in other samples from the cohort. When these RNA Rescue calls are added to the DNA Only calls, these genes achieve a statistically significant overall mutation rate for the cohort.

Here we specifically focus on results from 177 Endometrial Carcinoma (Kandoth *et al.*, 2013) and 230 Lung Adenocarcinoma (submitted) patients from TCGA. To demonstrate the increase in sensitivity from including the tumor RNA-Seq dataset, we simulated mutations by spiking them into real tumor DNA and tumor RNA using bamsurgeon (https://github.com/adamewing/bamsurgeon). We also evaluated sensitivity and specificity on real data using validation data that was generated by TCGA. We highlight RNA Rescue mutations found by the TBM in tumor suppressor genes such as *TP53*, *STK11*, and *CDKN2A* in Lung Adenocarcinoma.

## 2 Methods

RADIA operates on 2 or more BAM files, producing somatic mutation calls through a series of steps outlined in Fig. 1. Each step in this process is described in detail, beginning with the initial selection of sites for further processing and ending with a description of filters used to eliminate false positives while maintaining true positive calls.

### *2.1 Variant Detection with RADIA*

RADIA is typically run on 3 BAM (Li *et al.*, 2009) files consisting of a pair of patient-matched tumor and normal genomes and a tumor transcriptome and outputs germline (inherited) variants, Loss Of Heterozygosity (LOH) events, somatic Single Nucleotide Variants (SNVs), and RNA editing events. Here we focus specifically on the detection of somatic SNVs with RADIA. The DOM is run on the pairs of tumor and matched-normal DNA while the TBM is applied to the DNA and RNA triplets. After the DOM and TBM specific filters, the results are merged and run through a final read support filter (Fig. 1). If RNA-Seq data is not available, RADIA can utilize paired tumor and normal DNA genomes using the DOM to detect germline variants, LOH events, and somatic SNVs.

Internally, RADIA uses the samtools (Li *et al.*, 2009) mpileup command (version 0.1.18) to examine the pileups of bases in each sample in parallel. A heuristic algorithm determines the existence and type of variant at any given position based on the user-configurable minimum thresholds for overall depth, variant depth, Base Alignment Quality (BAQ) (Li, 2011), and mapping quality. Initially, RADIA requires a minimum overall depth of 4 bases, minimum variant depth of 2 bases, minimum phred BAQ of 10, and minimum phred mapping quality of 10. These initial calls are lenient in coverage and provide a good baseline set of calls for further filtering.

RADIA scans pileups of reads across the reference genome and outputs variants in Variant Call Format (VCF) (https://github.com/samtools/hts-specs). For each position, summary information such as the overall depth, allele specific depth and frequency, average BAQ base quality, average mapping quality, and the fraction of reads on the plus strand are calculated for both the DNA and RNA. All of this information is used during the filtering process.

## *2.2 Variant Filtering*

### 2.2.1 Filtering Around INDELs

Many current mutation calling algorithms have a pre-processing step to account for misaligned reads around INDELs. This realignment step is computationally expensive and relies on accurately predicting the location of INDELs which in itself is not a trivial problem. Base Alignment Quality (BAQ) is an alternative option for dealing with alignment ambiguity around INDELs. It calculates the probability that a base has been misaligned and returns the minimum of the original base quality and the base alignment quality. BAQ is run by default when executing a samtools mpileup command and has been shown to improve SNP calling accuracy (Li, 2011). We use the extended version of BAQ (option –E) that is activated by default in the latest version of samtools (0.1.19) for increased sensitivity and slightly lower specificity (Li *et al.*, 2009).

### 2.2.2 1000 Genomes Blacklist Filter

The 1000 Genomes Project coined the term “accessible genome” to be the part of the reference genome that is reliable for accurate variant calling after removing ambiguous or highly repetitive regions (1000 Genomes Project Consortium, 2010). Since the reference genome is incomplete, repetitive in places, and does not represent human genetic variation comprehensively, reads often get mapped incorrectly in locations outside the accessible genome (inaccessible sites), leading to false positive variant calls. Over 97% of inaccessible sites are due to high copy repeats or segmental duplications. In the pilot, the 1000 Genomes Project determined that 85% of the reference sequence and 93% of the coding region was accessible. Due to longer read lengths (75-100 bp) and improvements to both paired end protocols and sequence alignment algorithms, the accessible genome has increased to 94% of the reference and 98% of the coding region in Phase I (1000 Genomes Project Consortium, 2012). We filter variants that are not in the accessible genome using the 1000 Genomes Phase I mapping quality and depth blacklists (ftp://ftp-trace.ncbi.nih.gov/1000genomes/ftp/phase1/analysis_results/supporting/accessible_genome_masks/).

### 2.2.3 Strand-Bias Filter

It has recently been shown that variant allele reads that occur exclusively on one strand are largely associated with false positive calls (Larson *et al.*, 2012). In order to account for this technical artifact, we filter based on the variant allele strand bias. If we have at least 4 total reads supporting the variant allele, then we apply the strand bias filter if more than 90% of the reads are on the forward strand or more than 90% are on the reverse strand.

### 2.2.4 Filtering by mpileup Support

RADIA can be executed on patient-matched pairs of tumor and normal DNA samples using the DOM to identify germline variants and somatic mutations. We first compare the matched normal DNA to the human reference genome. We require the normal DNA to pass the mpileup support filters described in Table 1 for all germline variants.

If no germline variant is found, we compare the tumor DNA to the matched normal DNA and the reference to search for somatic mutations. We require the normal DNA and tumor DNA to pass the mpileup support filters shown in Table 1 for all somatic variants. To ensure that we have the power to detect a possible germline variant at this site, we require that the normal DNA depth is 10 or more.

| **Filter** | **Germline** | **Somatic** | |
|---|---|---|---|
| | Normal DNA | Normal DNA | Tumor DNA |
| Min Total Depth | 10 | 10 | 10 |
| Min Alt. Depth | 4 | NA | 4 |
| Min Alt. Percent | 10% | NA | 10% |
| Min Avg. Alt. BAQ | 20 | NA | 20 |
| Max Alt. Strand Bias | 90% | NA | 90% |
| Max Alt. Percent | NA | 2% | NA |
| Max Other Percent | 2% | 2% | 2% |

**Table 1. DNA Only Method mpileup support filters.** The germline variants and somatic mutations from the DOM are filtered according to the parameters described here. The minimum average alternative read BAQ filter uses the phred scale. The maximum other percent restricts the percentage of reads that are allowed to support an additional alternative allele.

We use the Triple BAM Method to augment our somatic mutation calls using both the pairs of DNA and the RNA-Seq data. The normal DNA, tumor DNA, and tumor RNA must pass the mpileup support filters shown in Table 2 for all somatic variants. We require at least one read with a minimum BAQ phred score of 15 in the tumor DNA to make a call. To rule out possible germline variants, we again require that the normal DNA depth is 10 or more. In addition, we filter out calls that overlap with dbSNP. We found that many false positive variants overlapped with earlier versions of dbSNP. These variants were due to technical artifacts and were removed from subsequent versions of dbSNP (Musumeci *et al.*, 2010). Therefore, we filter out all variants that overlap with dbSNP versions 130, 132 or 135 (ftp://ftp.ncbi.nih.gov/snp/). The TBM calls are subjected to further filtering procedures as shown in Fig. 1 and described below.

| **Filter** | **Somatic** | | |
|---|---|---|---|
| | Normal DNA | Tumor DNA | Tumor RNA |
| Min Total Depth | 10 | 1 | 10 |
| Min Alt. Depth | NA | 1 | 4 |
| Min Alt. Percent | NA | NA | 10% |
| Min Avg. Alt. BAQ | NA | 15 | 15 |
| Max Alt. Strand Bias | NA | 90% | 90% |
| Max Alt. Percent | 10% | NA | NA |
| Max Other Percent | 10% | 10% | 2% |

**Table 2. Triple BAM mpileup support filters.** The somatic mutations from the TBM are filtered according to the parameters shown here.

### 2.2.5 Pseudogene Filter

We noticed that many of our TBM mutations overlapped with predicted pseudogenes. Although expressed pseudogenes have recently been reported to be significant contributors to the transcriptional landscape and shown to play a role in cancer progression (Kalyana-Sundaram *et al.*, 2012), mutations that overlap with predicted pseudogenes have a high false positive rate. Sequence similarity of pseudogene copies to their parent genes leads to uncertainty in alignment within these regions. Because of these technical artifacts, we remove TBM mutations that overlap with pseudogenes (Baertsch *et al.*, 2008; Harrow *et al.*, 2012).

### 2.2.6 Highly Variable Genes Filter

We remove TBM mutation calls that overlap with families of genes that have high sequence similarity. Some examples of these gene families are Major Histocompatibility Complexes (MHCs), Human Leukocyte Antigens (HLAs), Ribosomal Proteins (RPLs), immunoglobulins and zinc fingers. While mutations in these genes may exist, special processing would be needed to distinguish them from false positive calls due to misaligned reads.

### 2.2.7 Positional Bias Filter

False positive calls are sometimes associated with misaligned reads where the alternative allele is consistently within a certain distance from the start or end of the read. The positional bias filter is applied when 95% or more of the reads that have an alternative allele are such that the alternate allele falls in the first third or last third of the read.

### 2.2.8 BLAT Filter

We observed multiple instances where RNA-Seq reads appeared to be incorrectly mapped due to the added difficulties in aligning RNA-Seq data, such as dealing with hard to identify splice junctions and multiple gene isoforms. To guarantee that the RNA-Seq reads that support a variant do not map better to another location in the genome, we created a BLAT filter. All of the RNA-Seq reads that support a variant were extracted from the BAM file and aligned to the human genome using BLAT (Kent, 2002). If the read mapped to another location with a better score, then the read was rejected. After using BLAT on each read, we again require that there are 4 valid reads that support the variant and that 10% or more of the reads support the variant.

### 2.2.9 Read Support Filter

We merge the calls from the DOM and the TBM and apply one final filter. We require that each somatic mutation be supported by at least 4 "perfect" reads. We define a perfect read as follows:

1. Minimum mapping quality of read is 10
2. Minimum base quality of alternative allele in read is 10
3. Minimum base qualities of the 5 bases up- and down-stream of the alternative base are 10
4. Read is properly paired
5. Read has less than 4 mismatches across its entirety when compared to the reference
6. Read doesn't require an insertion or deletion to be mapped

After determining the number of perfect reads that support the reference and the alternative at a coordinate, we re-apply the strand bias filter to guarantee that no more than 90% of the total perfect reads are from one strand.

# 3 Results

## 3.1 Sensitivity on Simulation Data

In order to evaluate sensitivity and demonstrate the increase in power from including the RNA sequence data, we simulated somatic mutation calls starting from real data. We spiked mutations into a pair of breast cancer tumor DNA and tumor RNA samples using bamsurgeon (https://github.com/adamewing/bamsurgeon), a tool we developed to generate simulation data that closely mimics actual experimental data from high-throughput sequencing datasets. Bamsurgeon first determines the loci that have an appropriate DNA and RNA depth to spike in mutations. It then extracts the reads at the loci, adjusts the VAF according to the user-defined VAF distribution, and then re-maps the reads (Supplementary Figure 2). This simulation strategy is more sophisticated than simply generating simulated reads from a reference genome, as it retains the biological and technical artifacts that are inherently present in next generation sequencing data.

We performed two spike-in experiments: one varying the DNA VAF while holding the RNA VAF constant, and inversely, varying the RNA VAF while holding the DNA VAF constant. To evaluate the sensitivity of RADIA, we spiked in 1,594 mutations to the tumor DNA sequence with a variant allele frequency ranging from 1-50% and to the tumor RNA sequence at a constant frequency of 25%. We used the original tumor DNA in place of the matched-normal and made mutation calls with RADIA.

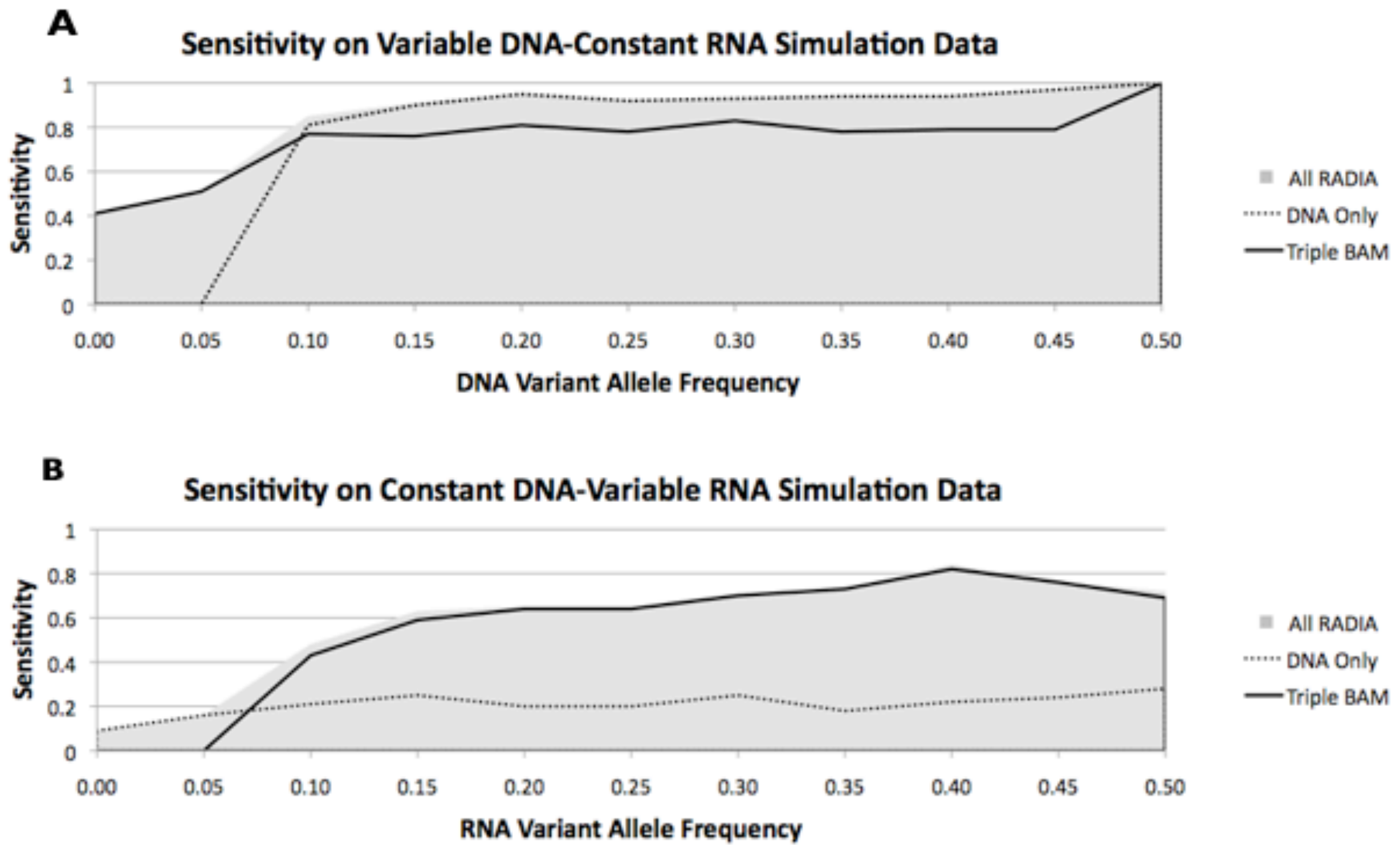

**Fig. 2 Sensitivity of RADIA on simulated data where artificial mutations were spiked into the tumor DNA and RNA of a breast cancer patient using bamsurgeon.** (A) Mutations were spiked into the DNA at VAFs from 1-50%. Mutations in the RNA were spiked in at a constant 25%. The overall sensitivity of RADIA was 86%. RNA Rescue calls from the Triple BAM method detected the mutations that had a DNA VAF < 10%. (B) Mutations were spiked into the DNA at 10% or less and into the RNA from 1-50%. Most of the DOM calls are filtered due to the low DNA allelic frequency. The calls that have adequate RNA read support are rescued back at these low DNA allelic frequencies.

The overall sensitivity rate averaged across all VAFs is 85% consisting of 1,351 out of 1,594 spiked in mutations (Fig. 2A). Of the 243 calls that were filtered out, over 50% are removed because they failed to meet the minimum variant allele frequency, more than 20% land in blacklist regions that the method ignores, and nearly 20% are discarded due to the BLAT filter. The number of mutations that are rejected by the full list of filters can be found in Supplementary Figure 3.

To demonstrate the ability of the TBM to rescue back calls at low DNA VAFs, we spiked in 1,761 mutations to the tumor RNA sequence with a variant allele frequency ranging from 1-50% and to the tumor DNA sequence at 10% or less. Most of the mutations by the DOM are filtered out due to the low allelic frequency in the DNA (Supplementary Figure 4). For the calls that have sufficient read support in the RNA, these low DNA VAFs are rescued back (Fig. 2B).

### *3.2 Specificity and Sensitivity on Real Data*

We made somatic mutation calls on 177 non-hypermutated TCGA endometrial carcinoma samples (Kandoth *et al.*, 2013). For these 177 samples, there exists validation data from the tumor and matched-normal DNA samples. We utilized a validation criterion similar to the one used by the TCGA Endometrial Working Group to validate the multi-center network mutation calls (Kandoth *et al.*, 2013). For each somatic mutation, we queried the patient-matched tumor and normal validation data. We required at least 10 reads in both the tumor and normal data in order to determine if a call validated, otherwise we classified it as ambiguous. If the variant was present at low levels in both datasets, we also classified it as ambiguous. Otherwise, we determined whether a mutation validated as germline/LOH, somatic, or neither according to Table 3. In addition, any RNA Rescue call in the "Not Validated" group that overlapped with COSMIC was considered as validated.

| Normal VAF | Tumor VAF | | | |
|---|---|---|---|---|
| | 0% | < 8% | ≥ 8%, < 20% | ≥ 20% |
| = 0% | Not Validated | Somatic Low | Somatic Med | Somatic High |
| < 3% | Not Validated | Ambiguous | Somatic Med | Somatic High |
| ≥ 3% | Germline/LOH | Germline/LOH | Germline/LOH | Germline/LOH |

**Table 3. Validation strategy in endometrial data.** Validation BAMs were used to determine the validation status for somatic mutations as shown here. A mutation is considered validated in the Somatic Low, Med, or High groups (blue) and not validated in the "Not Validated" (green) and Germline/LOH groups (red).

We made a total of 27,900 somatic mutation calls over 177 endometrial samples, of which 27,390 and 6,325 calls were made by the DOM and TBM respectively. Of the 6,325 TBM calls, there were 5,815 RNA Confirmation calls that were made by both the DOM and TBM signifying high DNA and RNA support, and importantly, a total of 510 RNA Rescue calls that were missed by the DOM.

Using the validation strategy described above, we demonstrate that the overall specificity for RADIA is 98% (Fig. 3A). Due to lack of coverage or uncertainty in the tumor and normal validation BAMs, a total of 1,825 calls were considered to be ambiguous. 25,520 mutations validated as somatic, 271 validated as germline/LOH variants and 284 were not validated. The specificity of calls made by the DOM and the TBM were 98% and 98.5% respectively. Over 99% of the calls that were made by both the DOM and the TBM were validated. There were 510 RNA Rescue calls that were made only by the TBM, and even though most of these calls were not targeted for validation, 81% of them validated as somatic. 175 of the 210 (83%) RNA Rescue calls validated using the validation BAMs while the remaining 35 (17%) were validated by COSMIC.

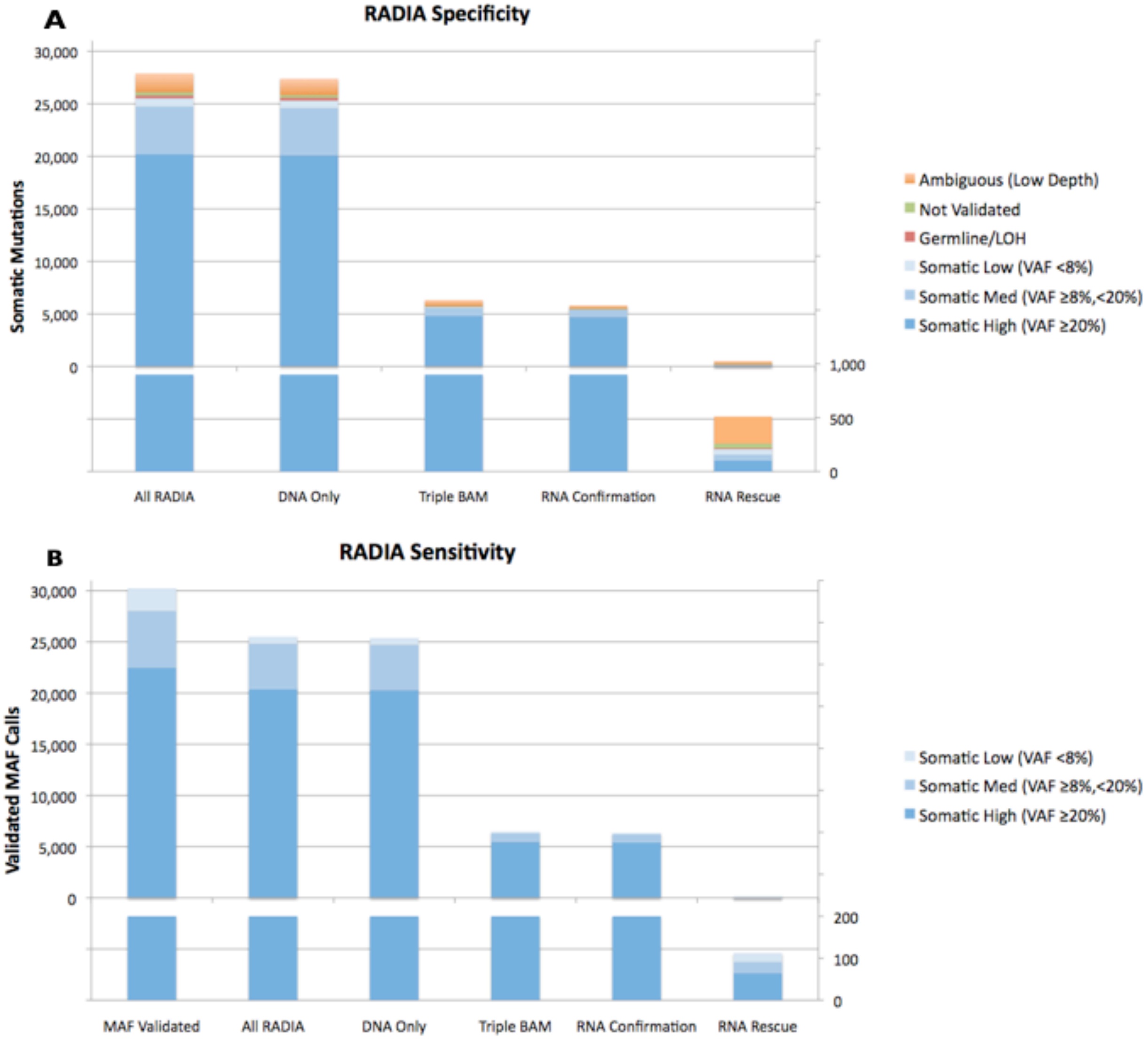

**Fig 3. Specificity and sensitivity of RADIA on 177 non-hypermutated endometrial carcinoma patients.** Mutations are considered validated in the Somatic Low, Med, or High groups (blue), not validated in the "Not Validated" (green) and Germline/LOH (red) groups, and Ambiguous (orange) when there was low read depth (<10 reads) or ambiguity in the validation data. (A) An overall specificity of 98% was demonstrated. RNA Confirmation calls with strong DNA and RNA support validated over 99%. RNA Rescue calls validated at 81%. (B) The union of all calls submitted by the TCGA network that validated as somatic was considered as the truth set. RADIA demonstrated an overall sensitivity rate of 84%. Of the calls that were missed, 33% occur at low variant allele frequencies and 23% occur in blacklist regions that are ignored.

In order to measure the sensitivity of RADIA, we considered the union of all calls submitted by the TCGA network from the Broad Institute and The Genome Institute at Washington University that validated as somatic as our truth set. There were 30,239 calls that validated as somatic from the TCGA network. We compared our somatic mutation calls to this truth set and demonstrated overall sensitivity of 84% (Fig. 3B, Supplementary Figure 5). Of the 4,751 calls that were missed, 1,539 (33%) were filtered by RADIA because they had a variant allele frequency less than 8% (Supplementary Figure 6). In addition, 1,072 (23%) landed in blacklist regions that were not considered (Supplementary Figure 6).

Finally, RADIA somatic mutation calls were analyzed during the course of our participation in the TCGA Lung Adenocarcinoma Working Group (submitted). Validation was performed by the Broad Institute on 74 genes of interest along with an additional 1,150 somatic SNVs. Validation was attempted on 2,404 RADIA calls and 2,395 (99.63%) were verified. From the DOM, 2,336 of the 2,345 calls (99.62%) validated. Importantly, 469/469 (100%) of the TBM calls consisting of 410 RNA Confirmation and 59 RNA Rescue calls were validated.

## *3.3 Somatic Mutations in Specific Genes in Lung Adenocarcinoma*

Mutations in the tumor suppressor gene *TP53* are common in the majority of human cancers. Most of the mutations occur in the DNA-binding domain (DBD) and are considered change-of-function mutations that alter activity of *TP53*, sometimes acting in a dominant negative manner to sequester wildtype tp53 protein *in trans* (Friedman *et al.*, 1993). As such, many p53 mutant proteins endow cells with oncogenic characteristics by promoting cell proliferation, survival, and metastasis (Muller and Vousden, 2012).

We ran RADIA on 230 TCGA Lung Adenocarcinoma triplets (submitted) and discovered two non-synonymous *TP53* mutations that were below the detection threshold for other analysis procedures used by the TCGA network working group (Table 4). Both of the mutations were found in COSMIC and reported in other lung cancer studies (Fouquet *et al.*, 2004; Liu *et al.*, 2002; Mori *et al.*, 2004; Pelosi *et al.*, 2012; The Cancer Genome Atlas, 2012). One of the mutations (G266E) is a loss-of-function mutation (Alsner *et al.*, 2000; Fernandez-Cuesta *et al.*, 2012; Pfaff *et al.*, 2010), while the other mutation (G199V) is an anti-apoptotic gain-of-function mutation that promotes cell survival through the signal transducer and activator of transcription-3 (STAT3) pathway (Kim *et al.*, 2009). Knockdown experiments of G199V p53 mutants demonstrated a level of anti-tumor activity similar to high doses of chemotherapeutic agents, suggesting that inhibition of G199V p53 mutants may be beneficial for cancer treatment (Kim *et al.*, 2009).

| Gene | Mutation | DNA VAF | RNA VAF |
|---|---|---|---|
| *TP53* | G266E | 1/7 (13%) | 6/10 (60%) |
| *TP53* | G199V | 4/64 (6%) | 8/57 (14%) |
| *STK11* | W239* | 1/13 (7%) | 20/40 (50%) |
| *CDKN2A* | R131H | 3/45 (7%) | 22/62 (35%) |
| *CDKN2A* | R122*/R163* | 2/16 (13%) | 31/34 (91%) |

**Table 4. RNA Rescue mutations in lung adenocarcinoma that were below the detection threshold for other mutation callers used by the TCGA network working group.** The ratio of reads supporting the calls along with the variant allele frequencies are shown for both the DNA and RNA.

Additionally, we found mutations in other well-known tumor suppressor genes such as *STK11* and *CDKN2A*. In the Lung Adenocarcinoma manuscript from TCGA, mutations in *STK11* and *CDKN2A* were reported in 17% and 4% of all patients, respectively (submitted). *STK11* was the fourth most mutated gene and *CDKN2A* was the sixteenth (submitted). The proximal-proliferative subtype in Lung Adenocarinoma is characterized by an enrichment of mutations in *KRAS* along with inactivation mutations in *STK11* (submitted). In the *STK11* gene, we discovered a nonsense mutation at W239* in the structurally conserved protein kinase domain that was below the detection threshold for other mutation algorithms used by the network. This mutation introduces an early stop codon in exon 5 (of 10) leading to a truncated protein. This site is in COSMIC and was previously reported to be part of a 398 nucleotide deletion in a lung cancer study (Davies *et al.*, 2005).

In the *CDKN2A* gene, we found one nonsense mutation at R122*, R163* and one missense mutation at R131H, R80H that were both found in COSMIC. *CDKN2A* is silenced in many CpG island methylator phenotype-high (CIMP-High) tumors by DNA methylation (submitted), but mutations and deletions in *CDKN2A* also result in loss of function. The nonsense mutation at R122*, R163* results in an early stop codon in exon 2 (of 3-4, isoform dependent) leading to a truncated protein. Previous lung cancer studies (Andujar *et al.*, 2010; Blons *et al.*, 2008; Imielinski *et al.*, 2012) have reported frameshifts and deletions at this site. The missense mutation at R131H was also found in colon cancer (The Cancer Genome Atlas, 2012), clear cell sarcoma (Takahira *et al.*, 2004), and chronic myeloid leukemia (Nagy *et al.*, 2003) and confirmed as somatic in biliary tract cancer (Ueki *et al.*, 2004).

# 4 Discussion

The identification of somatic mutations is a key step in characterizing the cancer genome. Until now, mutation calling algorithms have concentrated on comparing just the normal and tumor genomes within the same individual. In the past few years, it has become common to also sequence the tumor transcriptome using RNA-Seq technologies. We have developed a new method called RADIA that combines the normal DNA, tumor DNA, and tumor RNA from the same individual to increase the power to detect somatic mutations.

The accurate detection of somatic mutations is complicated by biological and technical artifacts such as tumor purity and subclonality, varying allele frequencies, sequencing depths, and copy-number variation. There is a trade-off between high sensitivity and high specificity, such that it is difficult to achieve both. By including an additional dataset, we are increasing our ability to reliably detect mutations, especially at low variant allele frequencies (Supplementary Figure 7) where the signal to noise ratio becomes unfavorable. By combining the three datasets, we are also able to confirm the expression of a mutation, providing more clues to its likely functional effect. Confirming mutations through RNA-Seq is also advantageous for large genomic studies such as TCGA in providing a means for weak validation for mutation calls without costly resequencing for validation (Supplementary Figure 8). We showed here that over 99% of the mutations with both strong DNA and RNA support validated in endometrial cancer, suggesting that if one is not using calls in clinical practice but rather estimating overall frequencies of specific mutations in a research cohort, the extreme expense in validating every call may not be warranted.

With RADIA, we are able to detect mutations in important cancer genes such as *TP53* that were previously not identified by other algorithms because the signal was lost in the noise. Somatic mutations are commonly used to group patients into subtypes that are critical for diagnosis and treatment of the disease. Our ability to rescue back calls for individual patients will assist in correctly identifying each patient's specific subtype and consequently their treatment options.

## Acknowledgements

We would like to thank Sofie Salama, J. Zachary Sanborn, Christopher Wilks, and Todd Lowe for helpful discussions and feedback on this manuscript.

*Funding*: This work was supported by the National Cancer Institute [U24CA143858 to A.R., S.M., A.E., and J.Z.; R01CA180778 to J.S.; and U24CA180951 to J.Z.]; the National Human Genome Research Institute [U01ES017154 to A.R.]; a gift from Edward Schulak [to A.E.]; and the Howard Hughes Medical Institute [to D.H.].

*Conflict of Interest*: none declared.

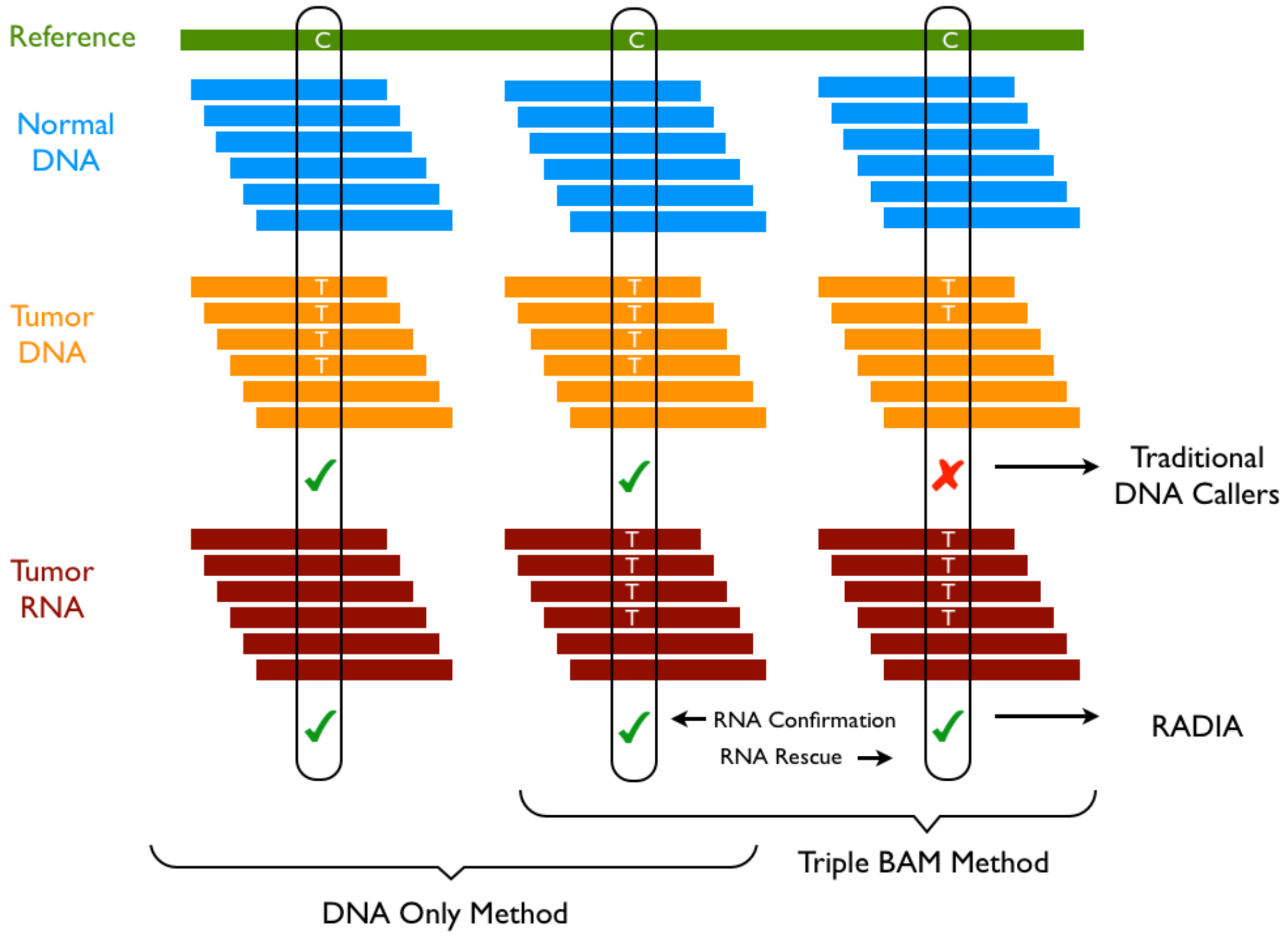

Supplementary Figure 1: Schematic of calls made by the DOM and TBM. In the first and middle columns, there is enough DNA read support for the DOM method and other algorithms acting on DNA pairs to make a call. In the middle and last columns, there is enough RNA read support for the TBM to make a call. The middle column illustrates "RNA Confirmation" calls that are detected by both the DOM and the TBM due to high read support in both the DNA and RNA. The last column represents the "RNA Rescue" calls that have some support in the DNA and strong evidence in the RNA.

| Cancer Type | Sample Count | Total Somatic SNVs | DNA Only Method | Triple BAM Method | RNA Rescue Calls | RNA Rescue Percent |
|---|---|---|---|---|---|---|
| Endometrial | 177 | 27900 | 27390 | 6325 | 510 | 2% |
| Lung Adenocarcinoma | 263 | 85044 | 79347 | 21484 | 5697 | 7% |
| Kidney Chromophobe | 66 | 4163 | 3957 | 1042 | 206 | 5% |
| Thyroid | 430 | 20849 | 19836 | 2882 | 1013 | 5% |
| Melanoma | 347 | 584431 | 573925 | 70091 | 10498 | 2% |
| Low-Grade Glioma | 289 | 13852 | 12837 | 3926 | 1015 | 4% |
| Prostate | 314 | 14630 | 12653 | 4631 | 846 | 6% |

Supplementary Table 1: Summary of TCGA samples analyzed. RADIA has been run on nearly 1,900 TCGA patients across seven different cancer types. The RNA Rescue calls make up 2-7% of the total somatic mutation calls across the seven types of cancer.

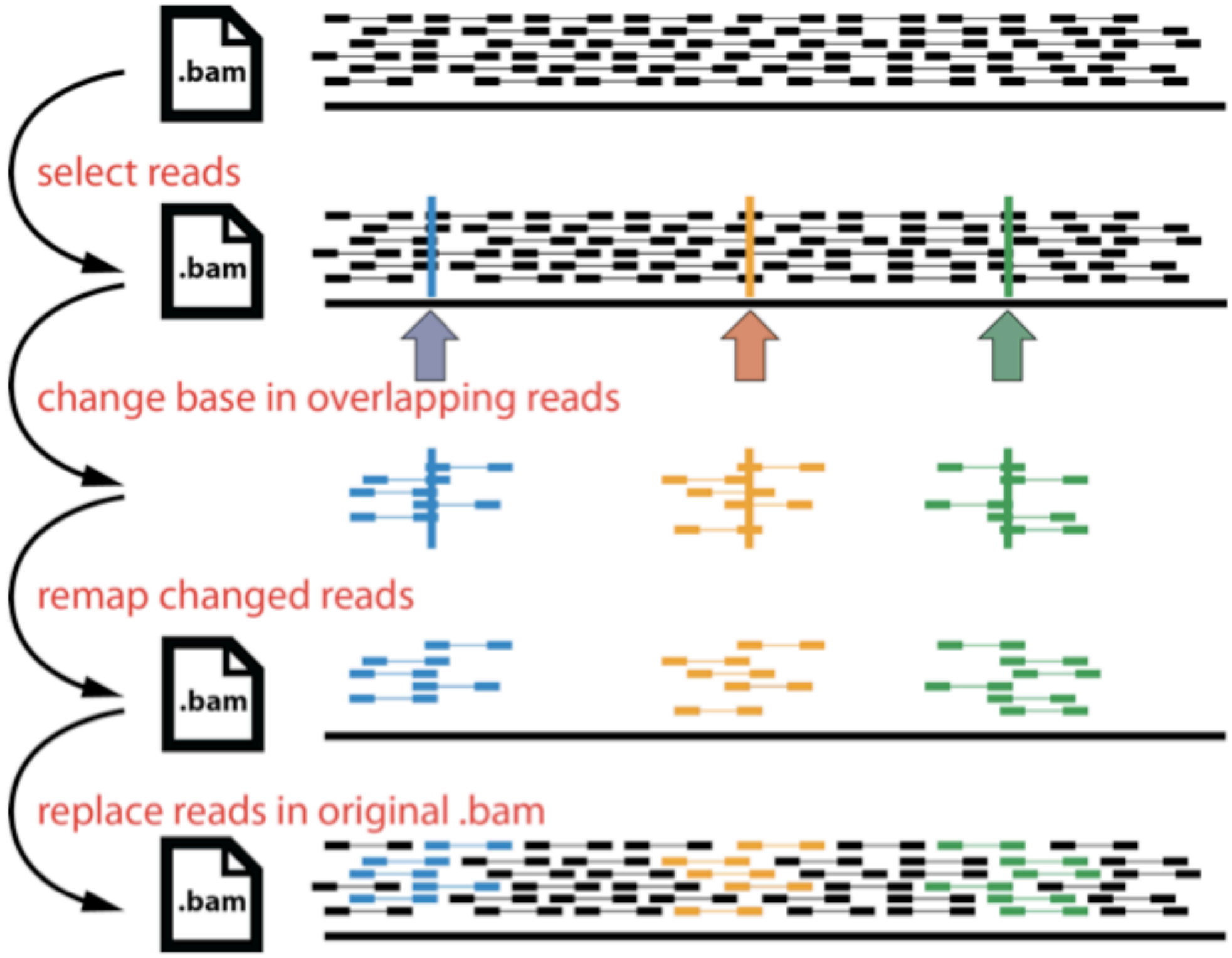

Supplementary Figure 2: Diagram of bamsurgeon methodology. Mutations are spiked into BAM files by selecting locations with adequate coverage and changing a number of the bases according to the desirable variant allele frequency distribution. Once the bases in the reads are changed, they are remapped to the genome, replacing the reads in the original BAM file.

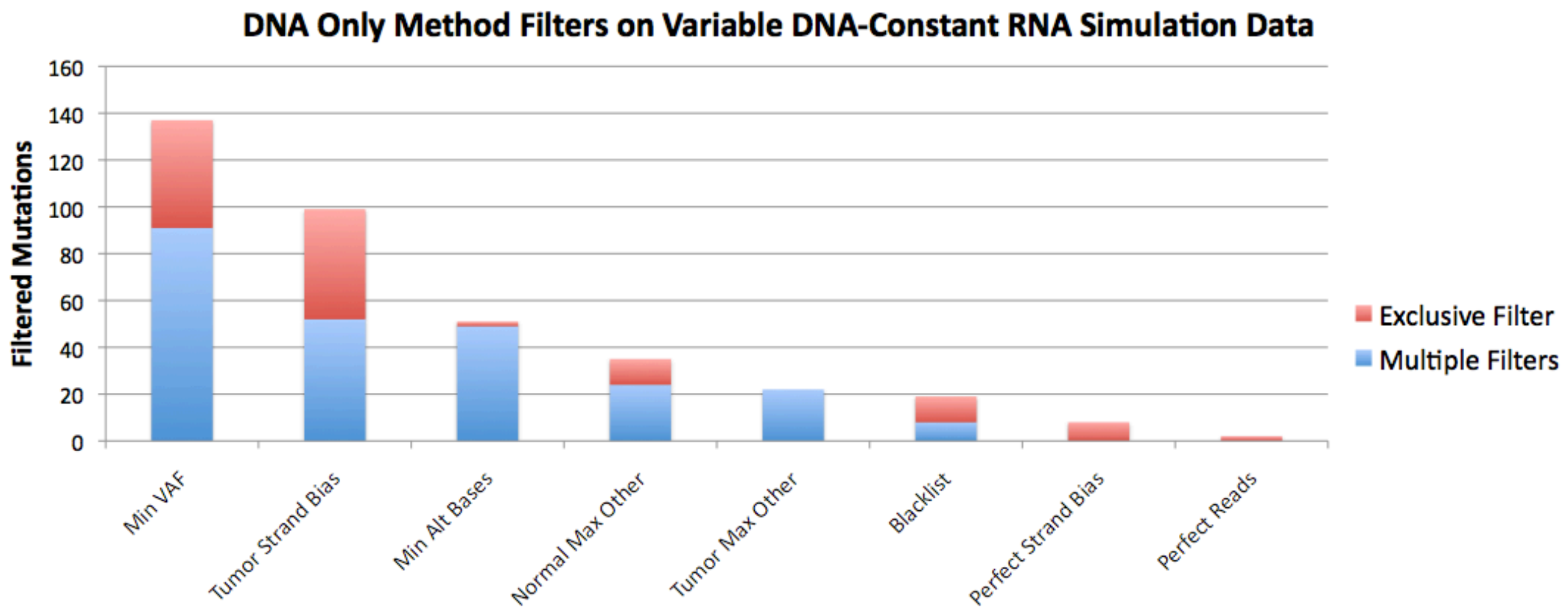

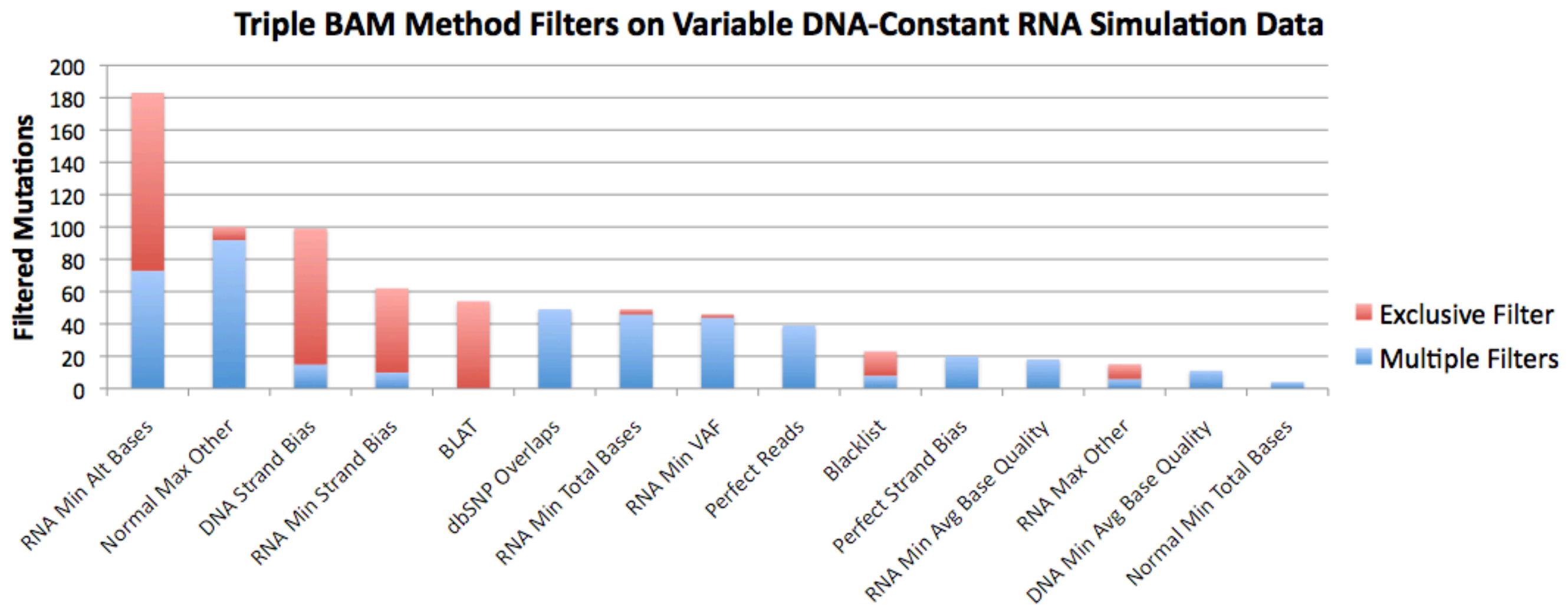

Supplementary Figure 3: Filters applied in the bamsurgeon simulation experiment where the DNA variant allele frequencies were distributed from 1-50% and the RNA was held constant at 25%. Most of the DOM calls were filtered because of the low variant allele frequency and tumor strand bias. In the TBM, most of the calls were filtered due to the minimum number of alternative alleles required to make a call (n=4) and strand bias in the tumor DNA and RNA.

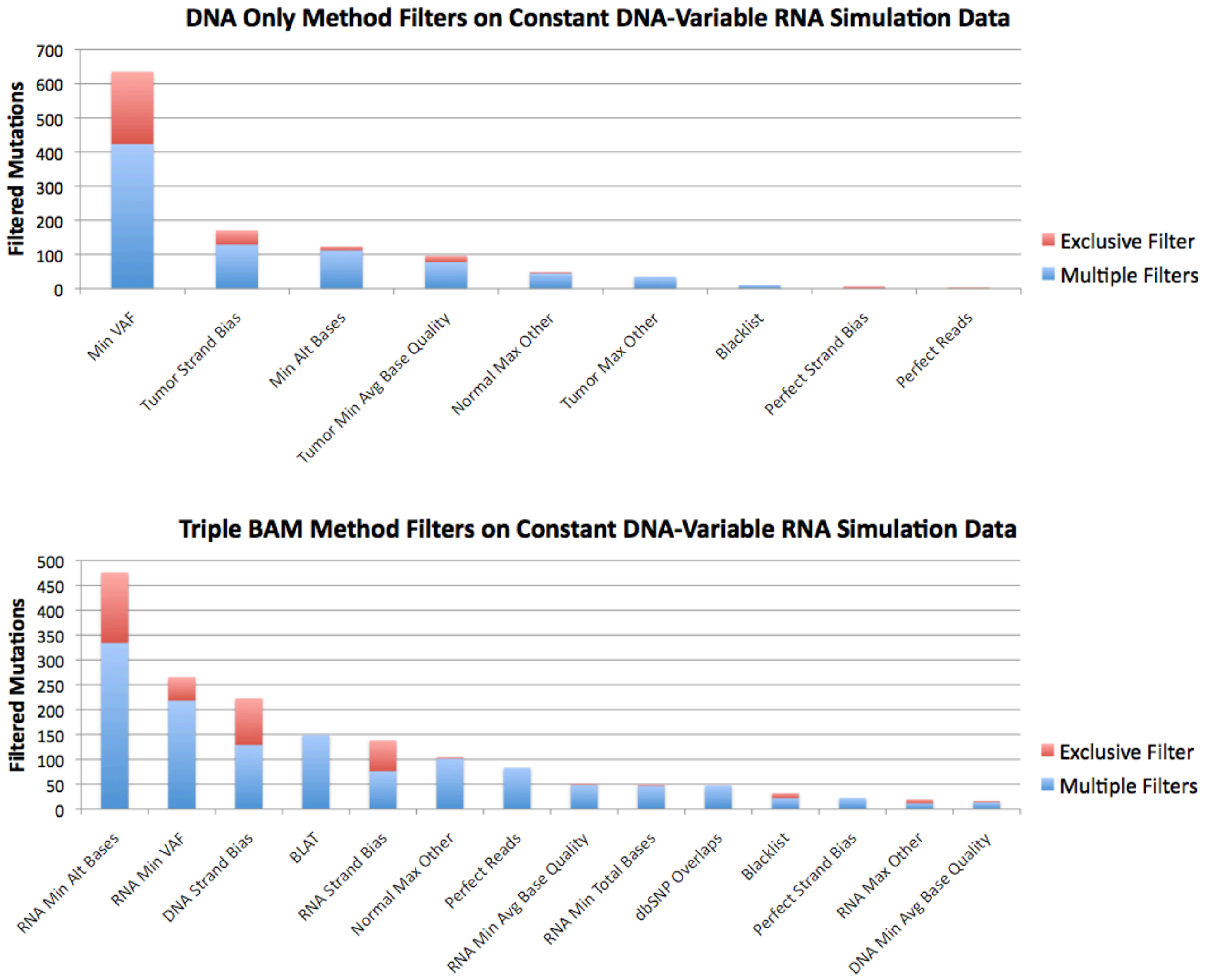

Supplementary Figure 4: Filters applied in the bamsurgeon simulation experiment where the RNA variant allele frequencies were distributed from 1-50% and the DNA was 10% or less. Most of the DOM calls were filtered because of the low DNA variant allele frequency and tumor strand bias. In the TBM, most of the calls were filtered due to the minimum number of alternative alleles required to make a call (n=4) and the low RNA variant allele frequency.

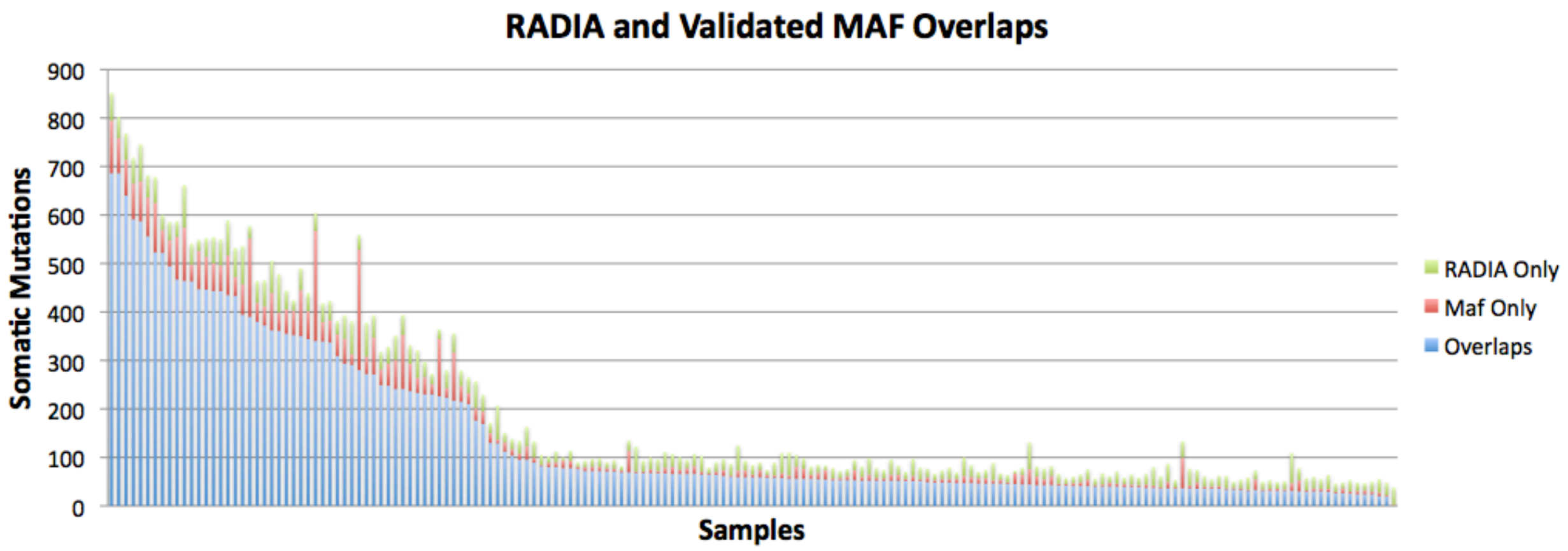

Supplementary Figure 5: The distribution of the overlaps between RADIA and the validated MAF calls.

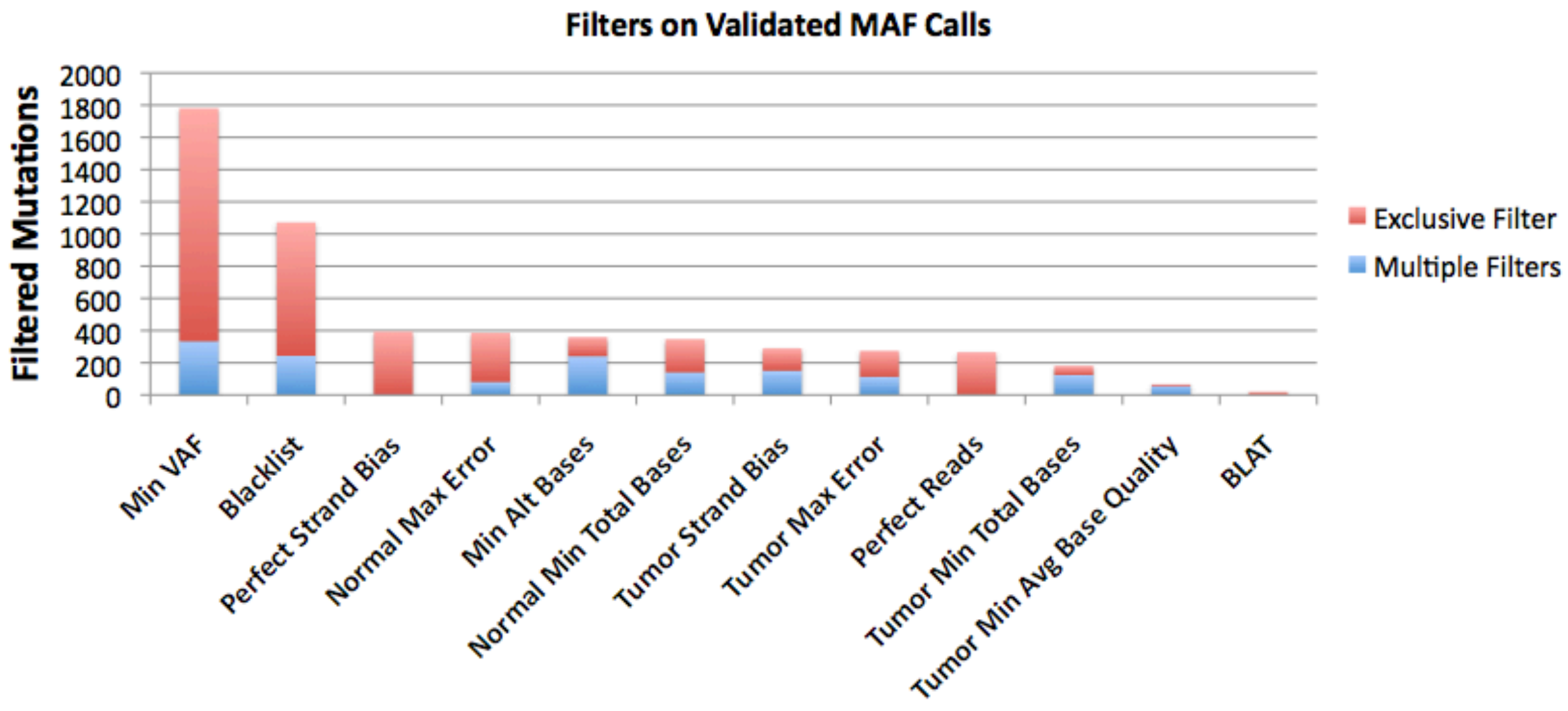

Supplementary Figure 6: Filters applied to the RADIA calls that validated as somatic in the MAF. 33% of the calls had a DNA VAF of 8% or less while 23% landed in blacklist regions that we didn't consider.

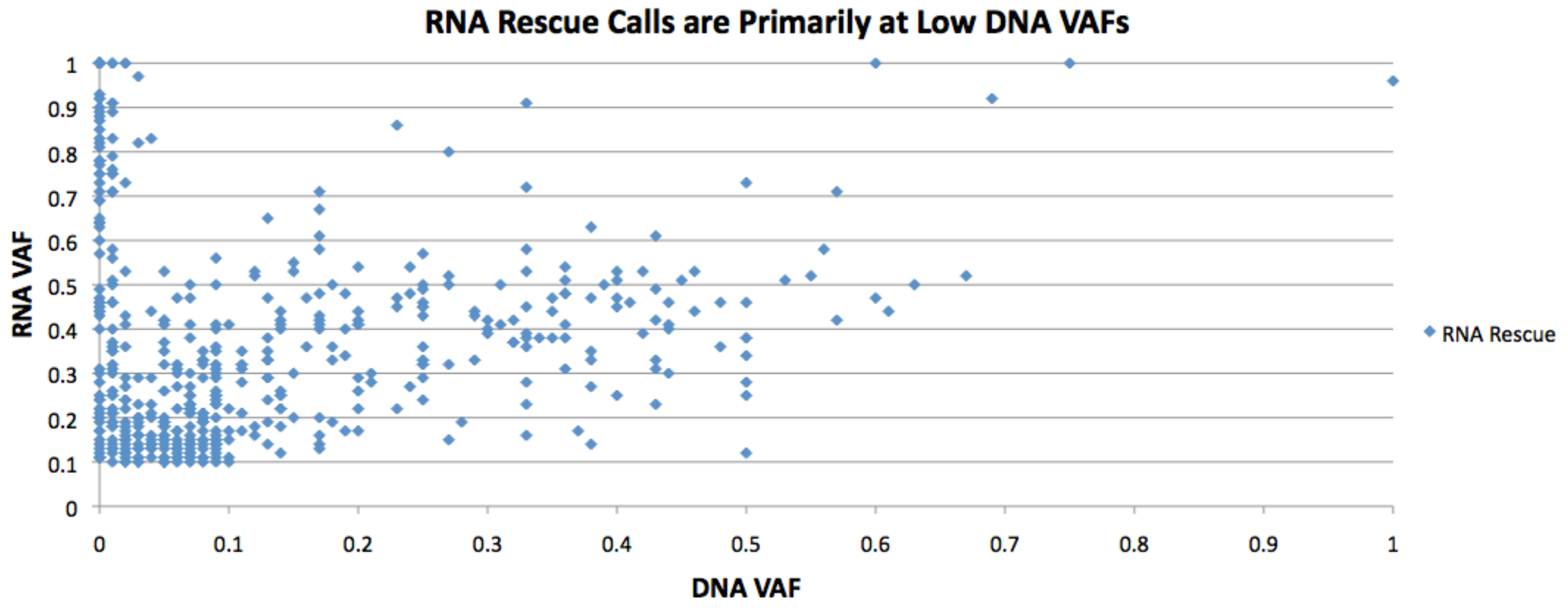

Supplementary Figure 7: RNA Rescue calls are primarily found at low DNA variant allele frequencies, but they are also able to rescue back calls at higher frequencies that were filtered due to non-depth related artifacts (e.g. strand-bias).

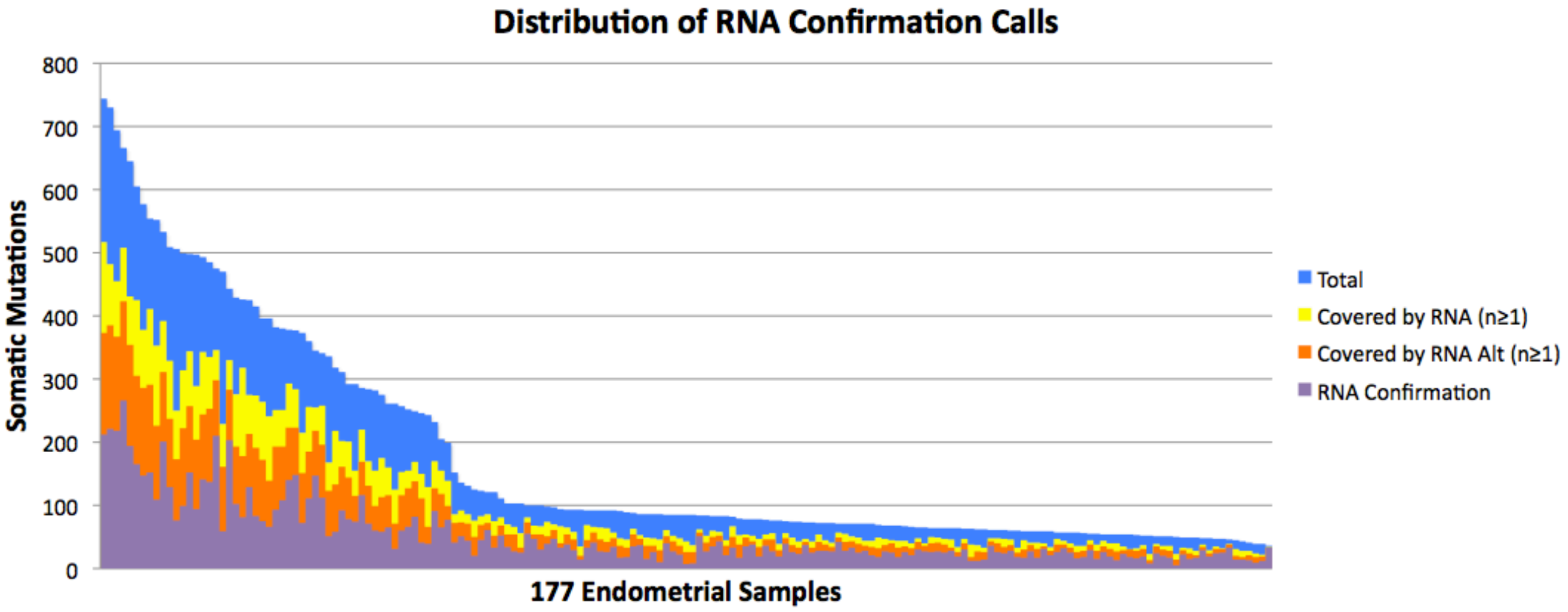

Supplementary Figure 8: The total number of mutations that are covered by at least one RNA read (yellow), one RNA read supporting the mutant allele (orange), and RNA Confirmation calls with high support in both the DNA and RNA (purple).